\documentclass[twocolumn]{aastex631} 

\usepackage{graphics,epsf}
\usepackage[utf8]{inputenc}
\usepackage{amsmath}                
\usepackage{amsfonts}               
\usepackage{amssymb}                
\usepackage{epsfig}                 
\usepackage{graphicx}               
\usepackage{float}
\usepackage{color}
\usepackage{multirow}               
\usepackage{array}
\usepackage{hyperref}
\usepackage{hhline}

\hypersetup{
    colorlinks=true,
    linkcolor=red,   
    urlcolor=cyan}

\usepackage[colorinlistoftodos]{todonotes}

\newcommand{\cm}{{~\rm cm}}
\newcommand{\km}{{~\rm km}}
\newcommand{\s}{{~\rm s}}

\newcommand{\g}{{~\rm g}}
\newcommand{\G}{{~\rm G}}
\newcommand{\K}{{~\rm K}}
\newcommand{\erg}{{~\rm erg}}

\begin{document}

\title{Simulating observed point-symmetric core-collapse supernova morphologies with the jittering jets explosion mechanism}
\date{December 2025}

\author[0009-0001-4877-1125]{Jessica Braudo}
\affiliation{Department of Physics, Technion - Israel Institute of Technology, Haifa, 3200003, Israel; jessicab@campus.technion.ac.il; amichaelis@campus.technion.ac.il; 
akashi@technion.ac.il;
soker@technion.ac.il}

\author[0000-0002-1361-9115]{Amir Michaelis}
\affiliation{Department of Mathematics and Physics, University of Haifa, Oranim Campus, Kiryat Tivon 3600600, Israel}
\affiliation{Department of Physics, Technion - Israel Institute of Technology, Haifa, 3200003, Israel; jessicab@campus.technion.ac.il; amichaelis@campus.technion.ac.il; 
akashi@technion.ac.il;
soker@technion.ac.il}

\author[0000-0001-7233-6871]{Muhammad Akashi} 
\affiliation{ Kinneret College on the Sea of Galilee, Samakh 15132, Israel}
\affiliation{Department of Physics, Technion - Israel Institute of Technology, Haifa, 3200003, Israel; jessicab@campus.technion.ac.il; amichaelis@campus.technion.ac.il; 
akashi@technion.ac.il;
soker@technion.ac.il}

\author[0000-0003-0375-8987]{Noam Soker}
\affiliation{Department of Physics, Technion - Israel Institute of Technology, Haifa, 3200003, Israel; jessicab@campus.technion.ac.il; amichaelis@campus.technion.ac.il; 
akashi@technion.ac.il;
soker@technion.ac.il}

\begin{abstract}
We conduct two three-dimensional hydrodynamic simulations of the jittering-jets explosion mechanism (JJEM) of core-collapse supernova (CCSN), launching three pairs of inclined opposite jets into the core of an enveloped-stripped stellar model, and reproduce some morphological features of observed CCSN remnants (CCSNRs) that a single pair of jets or instabilities alone cannot reproduce.   We launch the three pairs of jets within about a second, and follow the ejecta for more than 10 seconds until after shock breakout.
Our main findings are: (1) Although the jets are choked deep inside the star,  they manage to form a pronounced multipolar (point-symmetric) morphology. 
(2) Instabilities and vortices resulting from the jet-star interaction form small clumps and narrow filaments, some of which form point-symmetric morphology, resembling some observed CCSNRs.   (3) The most energetic jet of one simulation forms a large low-density blowout ahead of the ejecta, with filaments dragging behind it, resembling the blowout of the Cygnus Loop. (4) The inner ejecta presents two symmetry axes along two of the three jet axes: one of a pair of rings and one of a pair of nozzles, resembling the structure of the point-symmetric SNR J0450.4-7050. (5) The three pairs of jets compress two dense blocks between their axes. The blocks exhibit a Doppler-shift bipolar outflow highly inclined to the morphological axes along the jet axes. The inclined Doppler-bipolar outflow and morphology axis resembles the CCSNRe W49B and SNR G292.0+1.8.
Our study supports the claim that the JJEM is the primary explosion mechanism of CCSNe. 
\end{abstract}



\section{Introduction}
\label{sec:Introduction}

The collapsing core of core-collapse supernovae (CCSNe) releases a huge amount of gravitational energy as it forms a neutron star (NS; $\simeq 3 \times 10^{53} \erg$); neutrinos carry most of this energy, and a small fraction explodes the star ($\simeq 10^{50} - 10^{52} \erg$). Two alternative theoretical explosion mechanisms propose different processes for delivering a fraction of the gravitational energy of the collapsing core to power the explosion of the star: the jittering jets explosion mechanism (JJEM; for recent reviews, see \citealt{Soker2024UnivReview, Soker2025Learning}) and the neutrino-driven mechanism (delayed neutrino explosion mechanism; e.g., \citealt{Janka2025} and \citealt{Mezzacappa2026} for reviews). \cite{Soker2026G11} presents a correct account of the relation between the two explosion mechanisms and other energy sources, that either require an explosion before they operate, like the magnetar that, in most cases requires a jet-driven explosion (e.g., \citealt{Kumar2025}), and mechanisms that are very rare and at best account for a minority of CCSNe, like the magnetorotational explosion mechanism (e.g., \citealt{Shibataetal2025, Mannoetal2026, PanLi2026, Griffithsetal2026}) and the formation of a black hole in collapsars (e.g., \citealt{BoppGottlieb2025, Gottliebetal2025}, for recent papers).  

The point-symmetric type morphology of CCSN remnants (CCSNRs) is the observable that best distinguishes between the two theoretical explosion mechanisms, and can be decisive (e.g., \citealt{Soker2026G11}). According to the JJEM, the pairs of opposite jets that the newly born NS launches and explode the star, shape the ejecta to possess a point-symmetric morphology (e.g., \citealt{Braudoetal2025}), i.e., pairs of opposite (to the center) morphological features that do not share the same axis. In some CCSNRs, instabilities, interaction with an ambient medium, and other processes can smear the point-symmetric morphology (e.g., \citealt{SokerShishkin2025Vela}). Therefore, not all CCSNRs reveal point-symmetric morphologies, but many do. Opposite structural features can include rings, ears, lobes, nozzles, and more (e.g., \citealt{SokerAkashi2025, AkashiSoker2026a}). The identification in recent years of point-symmetric morphologies in about 20 CCSNRs, which is a large fraction of well-resolved CCSNRs that have not been distorted by the interstellar medium, constitutes a breakthrough in establishing the JJEM as the primary explosion mechanism of CCSNe (e.g., some papers since 2025, \citealt{BearSoker2025, Soker2025G0901, Soker2025N132D, Soker2025RCW89, Soker2026G11, Soker2025Dust,  Soker2026J0450, KlimovSoker2026, SokerShishkin2025Vela}). 
According to the JJEM, several to about twenty pairs of jets explode the star (e.g., \citealt{Soker2025Learning} for the parameters of the JJEM). Some CCSNe reveal several pairs of jets of about equal power, e.g., the Crab Nebula \citep{ShishkinSoker2025Crab}. 
Some CCSNRs reveal one, two, or three very powerful pairs of jets (e.g., SN 1987A; \citealt{Soker2026SN1987A}), and that one jet might be much more powerful than the opposite jet in a pair (e.g., \citealt{Bearetal2025Puppis, Shishkinetal2025S147}). These types of explosions can form two or three photospheric shells in the first several weeks of the explosion. The identifications of two (or three) photospheric shells in SN 2023ixf by \cite{SokerShiran2025} based on observation by \cite{Zimmermanetal2024} and by \cite{ShiranSoker2026} in SN 2024ggi based on observations by \cite{ChenTWetal2025}, constitute a strong support to the JJEM. 

While the identification of around 20 CCSNRs with point-symmetric and other jet-shaped structures strongly supports the JJEM, it severely challenges the neutrino-driven mechanism, and even rules it out as the primary explosion mechanism of CCSNe. For that, the tens of studies of the neutrino-driven mechanism in recent years ignore these observations and focus on simulating the revival of the stalled shock in stellar models, and comparing simulation results with some other observations (e.g., some papers from 2026: \citealt{
Akhmetalietal2026, ChenCHetal2026, EggenbergerAndersenetal2026, Giudicietal2026,  GogilashviliTamborra2026, GogilashviliTamborra2026b, JacobsonGalanetal2026, Kovalenkoetal2026, LuoZhaKajino2026, Murphyetal2026, Onoetal2026, Orlando2026, Paradisoetal2026, Rusakovetal2026, Shietal2026, VarmaMuller2026, Wessonetal2026, WuYetal2026}); the observations these study consider cannot distinguish between the two explosion mechanisms. 

Although the intensive study of jet-shaped point-symmetric morphologies and jet-shaped structures is only three years old in CCSNe, it is a mature field in the study of planetary nebulae  (e.g.,  \citealt{Morris1987, Soker1990AJ, SahaiTrauger1998, AkashiSoker2018,   EstrellaTrujilloetal2019, Tafoyaetal2019, Balicketal2020,   GarciaSeguraetal2020, GarciaSeguraetal2021, Clairmontetal2022, RechyGarciaetal2020, Danehkar2022, MoragaBaezetal2023, Ablimit2024, Derlopaetal2024, Mirandaetal2024, Sahaietal2024, Masaetal2026}), including precessing jets (e.g., \citealt{Guerreroetal1998, Mirandaetal1998, Sahaietal2005, Boffinetal2012, Sowickaetal2017, RechyGarciaetal2019, Guerreoetal2021, Clairmontetal2022}). \cite{Klimovetal2026} identified an S-shaped morphology in the enigmatic SNR 3C 397 which they attributed to a pair of precessing jets acting in the explosion. 
We here, as in the many other studies of the JJEM, apply the method that has been very successful in establishing the shaping processes of planetary nebulae. It includes a careful eye inspection and qualitative classification based on morphology (e.g., \citealt{Balick1987, Parkeretal2006, Sahaietal2007, Kwok2024}). Although qualitative, this method enables the robust identification of jet-shaped structures (e.g., \citealt{SahaiTrauger1998}), which motivates relevant numerical simulations, which in turn shed light on the shaping processes through qualitative comparisons of simulations with observations (e.g., \citealt{GarciaSeguraetal2021, GarciaSeguraetal2022, GarciaSeguraetal2025}) and by comparing planetary nebulae to CCSNRs (e.g., \citealt{Akashietal2018}).

We conduct simulations of multiple pairs of jets that interact with the core of an envelope-stripped star and explode it in the framework of the JJEM. We follow the jets into regions outside the core, namely, after they break out.  We differ from our previous study \citep{Braudoetal2025} by exploring additional jet properties and continuing the simulations after shock-breakout. We differ from simulations of wobbling jets that explode massive stars in rare cases (e.g., \citealt{Gottliebetal2022, Gottliebetal2023,BoppGottlieb2025}) by having jittering at large angles rather than the small-angle wobbling around a fixed axis, and by launching non-relativistic jets as a very common explosion process.  
We describe our numerical setup in Section \ref{sec:Numerical}, and the results in Sections \ref{sec:Results}, where we also compare them to four SNRs. We summarize in Section \ref{sec:Summary}.

\section{Numerical setup}
\label{sec:Numerical}

\subsection{Grid and initial conditions}
\label{subsec:Grid}

The numerical scheme is similar to the one we used in \cite{Braudoetal2025}; here, we describe the essential components.  
We perform three-dimensional (3D) hydrodynamical simulations with version 4.8 of the {\sc flash} code \citep{Fryxell2000}. The calculations use the unsplit hydrodynamics solver \citep{Lee2006, LeeDeane2009,Lee2013}. 
We simulate in full 3D on a Cartesian $(x,y,z)$ domain extending to $\pm 6\times10^{10} \cm$ in each coordinate direction, corresponding to a total box size of $1.2\times10^{11} \cm$ per axis; the stellar model center is at the origin. We impose outflow boundary conditions on all outer boundaries. 
We employ adaptive mesh refinement (AMR) with a maximum refinement level of 6 throughout the domain, while enforcing a higher effective resolution in the central region. Specifically, within a spherical radius of $5\times10^{8} \cm$ from the origin, the grid has AMR level 10, to better resolve the jet injection region and the early interaction of the jets with the collapsing core material.

We launch the jets into an initially spherically symmetric model of a collapsing massive stellar core. In the innermost region of the domain, we adopt the collapsing core density profile from \citet{PapishSoker2014Planar}, originally based on the post-bounce structure of a $15\,M_\odot$ progenitor at $t \simeq 0.2 \s$ after bounce \citep{Liebendorferetal2005}. This inner profile extends to radii of $2\times10^{9} \cm$ ($2\times 10^4 \km$), with densities ranging from $\rho \simeq 4\times10^{14} \g \cm^{-3}$ near the center down to $\simeq 6\times10^{4} \g \cm^{-3}$ at its outer boundary. 
Between $2\times10^{9} \cm$ and the stellar surface at $8\times10^{9} \cm$ ($8 \times 10^4 \km$), we attach an envelope structure of a hydrogen- and helium-stripped $15\,M_\odot$ Wolf–Rayet progenitor obtained from a MESA simulation. 
Across this extended stellar region, the density decreases to $\simeq 30 \g \cm^{-3}$ near the stellar surface.

Outside the star ($r > 8\times10^{9} \cm$), we impose a circumstellar material profile that continues to decline outward, reaching densities of $ 7\times10^{-5} \g \cm^{-3}$ at the edge of the computational domain. 
This extended setup allows us to follow the jet propagation from deep inside the collapsing core, through the outer stellar layers, and into the surrounding medium after breakout.

We model the gravitational field of the newly formed NS by imposing a central gravitational potential of an $1.4\,M_\odot$ point-mass. 
This central gravity governs the motion of the surrounding gas and plays a key role in shaping the fallback and backflow that might develop during the jet propagation.

For numerical efficiency and to allow multiple high-resolution simulations within the extended computational domain, the jets are injected at a radius of $5\times10^{7} \cm = 500 \km$ rather than at the physical launching scale of tens of kilometers. We include no additional NS wind or late-time jet activity that late fallback accretion might power. Consequently, we do not focus on the innermost region where disk formation and jet launching would physically occur, but instead follow the large-scale hydrodynamical evolution of the injected jets as they interact with the collapsing core and outer stellar layers.

The numerical code includes the electron-positron Helmholtz free energy Equation of State (EOS) \citep{TimmesSwesty2000, TimmesArnett1999, Aparicio1998}, with the temperature and density as the free variables. This EOS employs the Fermi-Dirac distribution for a noninteracting Fermi gas of ionized matter. It coincides with the gamma EOS at low densities and temperatures, while also accounting for denser matter where degeneracy plays a significant role, and incorporates Coulomb corrections and blackbody radiation pressure.

\subsection{The jittering jet pairs}
\label{subsec:Jets}

We launch, i.e., numerically inject, each jet within a conical region whose inner boundary is located at $r_{\rm in,j}=5\times10^{7} \cm$ ($500 \km$), its outer boundary is at $r_{\rm out,j}=3\times10^{8} \cm$ ($3000 \km$), and whose half-opening angle is $\alpha_{\rm j}$. 
The jet's axis is specified by the angles $(\theta_{\rm j},\phi_{\rm j})$, where $\theta_{\rm j}$ denotes the polar angle measured from the $z$-axis and $\phi_{\rm j}$ is the azimuthal angle measured in the $(x,y)$ plane from the $x$-axis.
We launch the jets at the same initial velocity $v_{\rm j}=5\times10^{4}\ \mathrm{km\ s^{-1}}$, about twice the escape speed near the inner jet-injection radius. Each with an initial Mach number of 6.5. 
We launch the jets as opposite pairs, but not in all cases are the two opposite jets equal in their properties. 

\subsection{Simulated cases}
\label{subsec:Cases}

According to the JJEM, the newly formed NS launches the jets from radii of $\simeq 20 - 50 \km$, and the jets propagate through the turbulent accretion flow below and above the stalled shock, mainly the zone extending $\simeq 50 – 200 \km$. The stochastic and intermittent nature of the accretion disk that launches the jets can lead to opposite jets in a pair (e.g., \citealt{Soker2024N63A}). To examine this effect, we perform two simulations: in one case, termed E3, the two jets in each pair are identical, while in the second case, termed D3, the opposite jets differ in their injected properties. 
Both simulations include three jet-launching episodes (three pairs), as we described in Tables~\ref{Tab:Table1} and~\ref{Tab:Table2}.
\begin{table}[h]
\scriptsize
\begin{center}
\caption{Initial parameters of simulation D3}
\begin{tabular}{|l|cc|cc|cc|}
\hline
Jets' pair & \multicolumn{2}{c|}{1} & \multicolumn{2}{c|}{2} & \multicolumn{2}{c|}{3} \\
\hline
Jet & + & - & + & - & + & - \\
\hline
$t_{\rm j}({\rm start})\ [\rm s]$ & $0.00$ & $0.00$ & $0.1$ & $0.1$ & $0.2$ & $0.2$\\
\hline
$\Delta t_{\rm j} {\rm(active)}\ [\rm s]$ & $0.1$ & $0.1$ & $0.07$ & $0.07$ & $0.05$ & $0.1$\\
\hline
$\theta_{\rm j}\ [\rm deg]$ & $35$ & $145$ & $0$ & $180$ & $90$ & $90$ \\
\hline
$\phi_{\rm j}\ [\rm deg]$ & $0$ & $180$ & $0$ & $0$ & $90$ & $270$ \\
\hline
Axis $(x,y,z)$      & \multicolumn{2}{c|}{$(\pm 0.574, 0, \pm 0.819)$} &  \multicolumn{2}{c|}{$(0,0, \pm 1)$}&  \multicolumn{2}{c|}{$(0, \pm 1 ,0)$}\\
\hline
$\alpha_{\rm j}\ [\rm deg]$ & $10$ & $10$ & $10$ & $20$ & $10$ & $20$\\
\hline
$\dot{m}_{\rm j}\ [10^{32} \g \s^{-1}]$ & $1$ & $3$ & $2$ & $2$ & $2$ & $2$ \\
\hline
$E_{\rm k, j}\ [10^{50} \erg]$ & $1.25$ & $3.75$ & $1.75$ & $1.75$ & $1.25$ & $2.5$ \\
\hline

\end{tabular}
  \\
\label{Tab:Table1}
\end{center}
\begin{flushleft}
\small 
Notes: Initial parameters of simulation D3, which consists of three unequal-jet pairs. All jets have an initial velocity of $v_{\rm j} = 50,000 \km \s^{-1}$, amounts to a Mach number of $\simeq 6.5$. The second row indicates the two opposite jets within each pair, denoted by “+” and “-”. The following rows list the starting time $t_{\rm j}({\rm start})$ and the active duration $\Delta t_{\rm j}({\rm active})$ of each individual jet. The three jet-launching episodes are separated by quiet intervals. The direction of each jet is $(\theta_{\rm j}, \phi_{\rm j})$, where $\theta_{\rm j}$ is the polar angle measured from the $z$-axis, and $\phi_{\rm j}$ is the azimuthal angle measured in the $xy$-plane relative to the $x$-axis. We also mark the directions of the two jets in the coordinates $(x,y,z)$. The parameter $\alpha_{\rm j}$ denotes the half-opening angle of each jet, $\dot{m}_{\rm j}$ is the mass injection rate of a single jet, and $E_{\rm k,j}$ is the kinetic energy carried by one jet. 
The jets in a pair might differ in mass injection rate, opening angle, duration, and kinetic energy. The total kinetic energy injected by all six jets in simulation D3 is $E_{\rm k,D3} =  1.225 \times 10^{51}\ \rm erg$.
\end{flushleft}
\end{table}
\begin{table}[h]
\scriptsize
\begin{center}
  \caption{Initial parameters of simulation E3}
\begin{tabular}{|l|cc|cc|cc|}
\hline
Jets' pair & \multicolumn{2}{c|}{1} & \multicolumn{2}{c|}{2} & \multicolumn{2}{c|}{3} \\
\hline
Jet         & + & - & + & - & + & - \\
\hline
$t_{\rm j}({\rm start})\ [\rm s]$ & $0.00$ & $0.00$ & $0.48$ & $0.48$ & $0.96$ & $0.96$\\
\hline
$\Delta t_{\rm j} {\rm(active)}\ [\rm s]$ & $0.08$ & $0.08$ & $0.08$ & $0.08$ & $0.08$ & $0.08$\\
\hline
$\theta_{\rm j}\ [\rm deg]$ & $35$ & $145$ & $0$ & $180$ & $90$ & $90$ \\
\hline
$\phi_{\rm j}\ [\rm deg]$ & $0$ & $180$ & $0$ & $0$ & $90$ & $270$ \\
\hline
Axis $(x,y,z)$      & \multicolumn{2}{c|}{$(\pm 0.574, 0, \pm 0.819)$} &  \multicolumn{2}{c|}{$(0,0, \pm 1)$}&  \multicolumn{2}{c|}{$(0, \pm 1 ,0)$}\\
\hline
$\alpha_{\rm j}\ [\rm deg]$ & $10$ & $10$ & $10$ & $10$ & $10$ & $10$\\
\hline
$\dot{m}_{\rm j}\ [10^{32} \g \s^{-1}]$ & $2$ & $2$ & $2$ & $2$ & $2$ & $2$ \\
\hline
$E_{\rm k, j}\ [10^{50} \erg]$ & $2$ & $2$ & $2$ & $2$ & $2$ & $2$ \\
\hline

\end{tabular}
  \\
\label{Tab:Table2}
\end{center}
\begin{flushleft}
\small 
Notes: Similar quantities and notation as in Table \ref{Tab:Table1}, but for simulation E3, which consists of three equal-jet pairs of jets; the jets differ only in their direction. 
The total kinetic energy injected by all six jets in E3 is 
$E_{\rm k,E3} = 1.2 \times 10^{51}\ \rm erg$.
\end{flushleft}
\end{table}

In simulation D3 (Table \ref{Tab:Table1}), each jet-launching episode consists of two exactly opposite in direction jets that differ in mass injection rate, hence kinetic energy (pair 1), opening angle (pairs 2 and 3), or duration, hence kinetic energy (pair 3). The temporal separation between consecutive jet-launching episodes is much shorter than in simulation E3. The total kinetic energy injected by the six jets in D3 is $E_{\rm k,D3} = 1.225\times10^{51}\ \rm erg$. The launching velocity of all jets is $v_{\rm j} = 50,000 \km \s^{-1}$ and their Mach number is $\simeq 6.5$.
In simulation E3 all jets are identical, implying equal-jet pairs (Table \ref{Tab:Table2}). The jet velocity is the same as in simulation D3, and the total kinetic energy of the jets is very similar, $E_{\rm k,E3} = 1.2\times10^{51}\ \rm erg$. As said, the total activity period of the jets is much longer than in simulation D3 .


\section{Point-symmetrical morphologies}
\label{sec:Results}

Aiming to motivate and facilitate comparison of our results with observed CCSNR morphologies, we discuss in turn the different simulated morphological features that both allow comparison with observations and help decide between the two theoretical explosion mechanisms, the JJEM and the neutrino-driven mechanism (Section \ref{sec:Introduction}). Namely, we will focus on morphological features we obtain from the JJEM simulations that the theoretical neutrino-driven mechanism cannot explain or predict to be rare.

\subsection{Multipolar point-symmetric morphologies}
\label{subsec:Multipolar}

\begin{figure*}
\begin{center}
\vspace{-0.0cm}
\includegraphics[trim=0.0cm 0.0cm 0.3cm 0.0cm ,clip, angle=0, scale=0.70]{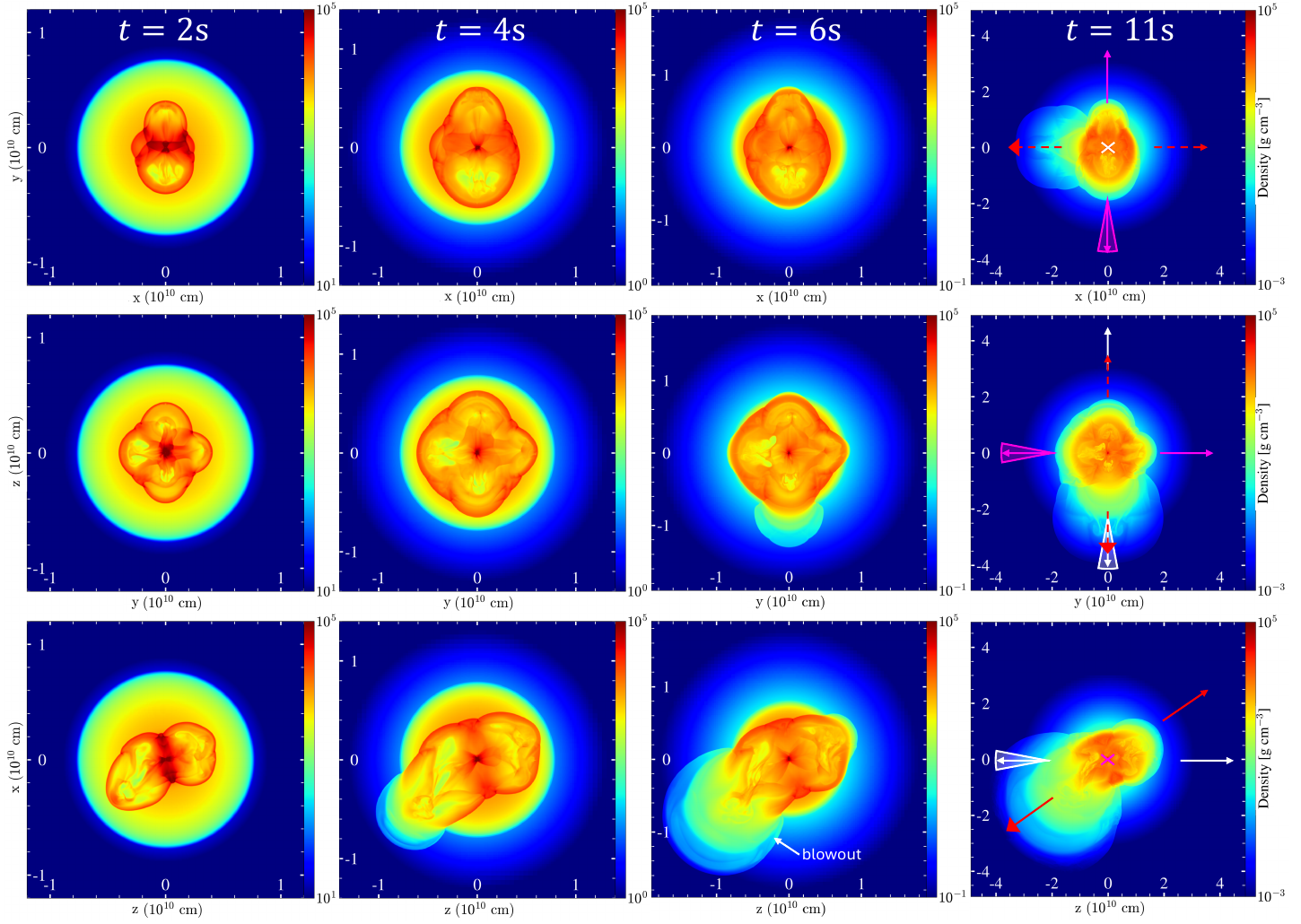} 
\vspace{-0.2cm}
\caption{Density maps of simulation D3 at three planes, from top to bottom, $z=0$, $x=0$, and $y=0$, at four times as indicated, long after the jet activity ceased. The density scale changes with time according to the color bars, in the ranges of, from deep blue to deep red: 
$10 - 10^5 \g \cm^{-3}$ at $t= 2 \s$; 
$1 - 10^5 \g \cm^{-3}$ at $t= 4 \s$;
$0.1 - 10^5 \g \cm^{-3}$ at $t= 6 \s$;
$0.001 - 10^5 \g \cm^{-3}$ at $t= 11 \s$. 
The arrows in the right panels are projections of the jets onto that plane (dashed lines), while solid arrows indicate that the jets are in that plane. A cross represents a pair of jets perpendicular to the plane.
Red color is for the first pair of jets, white for the second, and pink is for the third.  Big arrow heads correspond to a higher mass loss rate $\dot{m}_{\rm j}$, and a triangle corresponds to a wider half-opening angle $\alpha_{\rm j}$.
}
\label{Fig:densD3}
\end{center}
\end{figure*}

\begin{figure*}
\begin{center}  
\vspace{-0.0cm}
\includegraphics[trim=0.0cm 0.0cm 0.0cm 0.0cm ,clip, angle=0, scale=0.70]{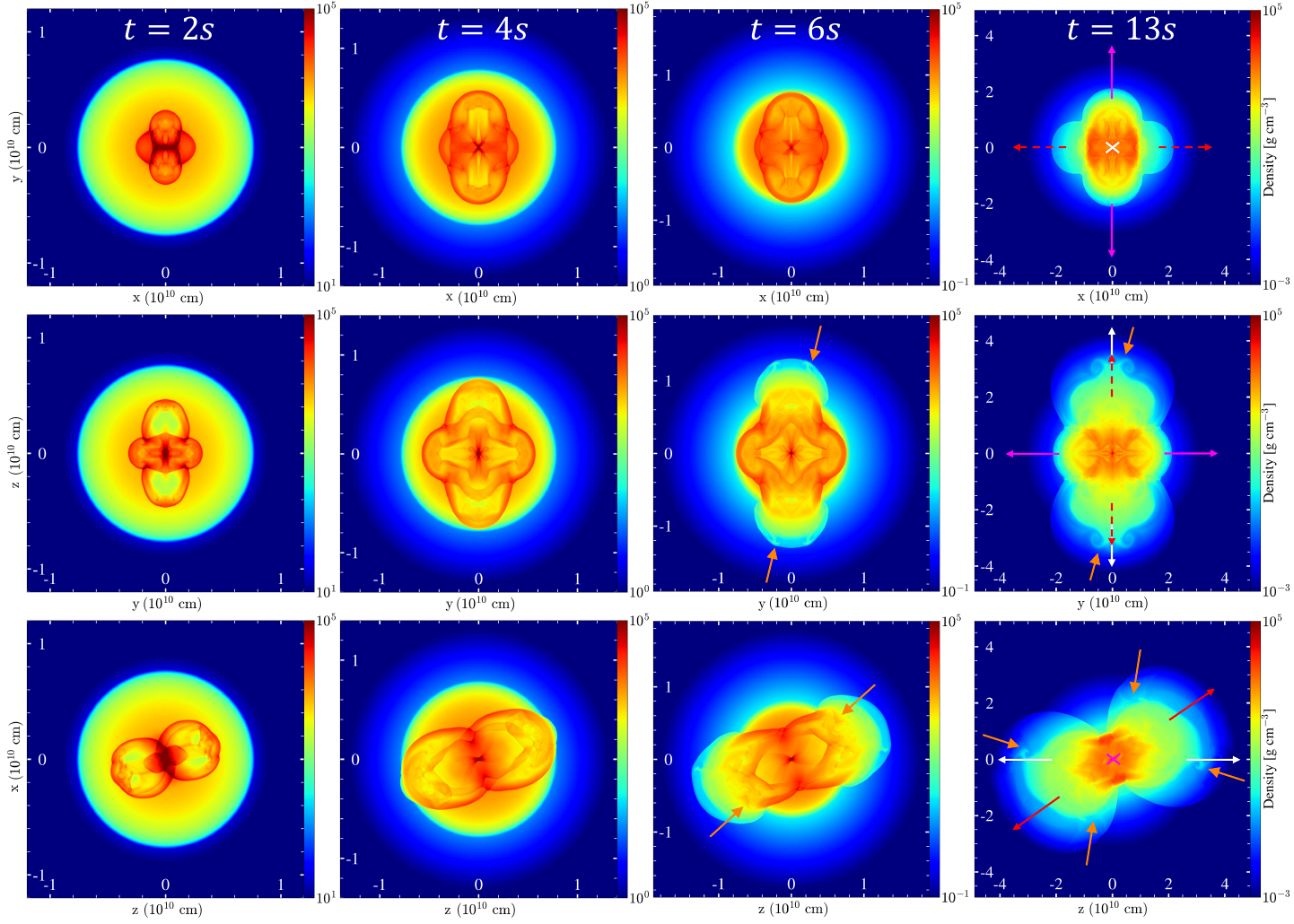} 
\vspace{-0.2cm}
\caption{Similar to figure \ref{Fig:densD3} but for simulation E3 with latest time $t=13\s$. Note that in this simulation, all jets are equal. The pairs of orange arrows pointing inward point at some pairs of clumps.
}
\label{Fig:densE3}
\end{center}
\end{figure*}

We present the density maps for simulation D3 in Figure \ref{Fig:densD3} and for simulation E3 in Figure \ref{Fig:densE3}, in the three grid-planes, $x=0$, $y=0$, and $z=0$, from upper to lower planes, and at four times, from left to right. To reveal both the outer low-density and inner high-density regions, we use logarithmic density scales, as shown in the vertical color bar to the right of each panel.
As the ejecta expands, the area the panels cover increases over time (from left to right). 

In both the different-jets (D3) and the equal-jets (E3) simulations, the pairs of jets inflate pairs of bubbles, as commonly observed in cooling flow clusters where jets are directly observed (for the comparison of bubbles in the hot gas in clusters of galaxies and of CCSNRs see \citealt{Soker2024CF}). The bubbles start to break out from the stellar surface, shock breakout,  at around $t \simeq 4 \s$. The three pairs of jets form three pairs of lobes, i.e., a multipolar morphology. Specifically, in the last column of Figures \ref{Fig:densD3} and \ref{Fig:densE3}, we identify protrusions along the three jet axes. The multipolar morphology is prominent in simulation E3 (Figure \ref{Fig:densE3}) where the jets are equal. In the last column of Figure \ref{Fig:densE3}, we clearly identify a bipolar elongation along each of the three jet axes; we mark the three jet axes on these panels.  
   
\begin{figure}
\begin{center}
\vspace{-0.0cm}
\includegraphics[trim=2.0cm 2.0cm 3.0cm 2.0cm ,clip, angle=270, scale=0.33]{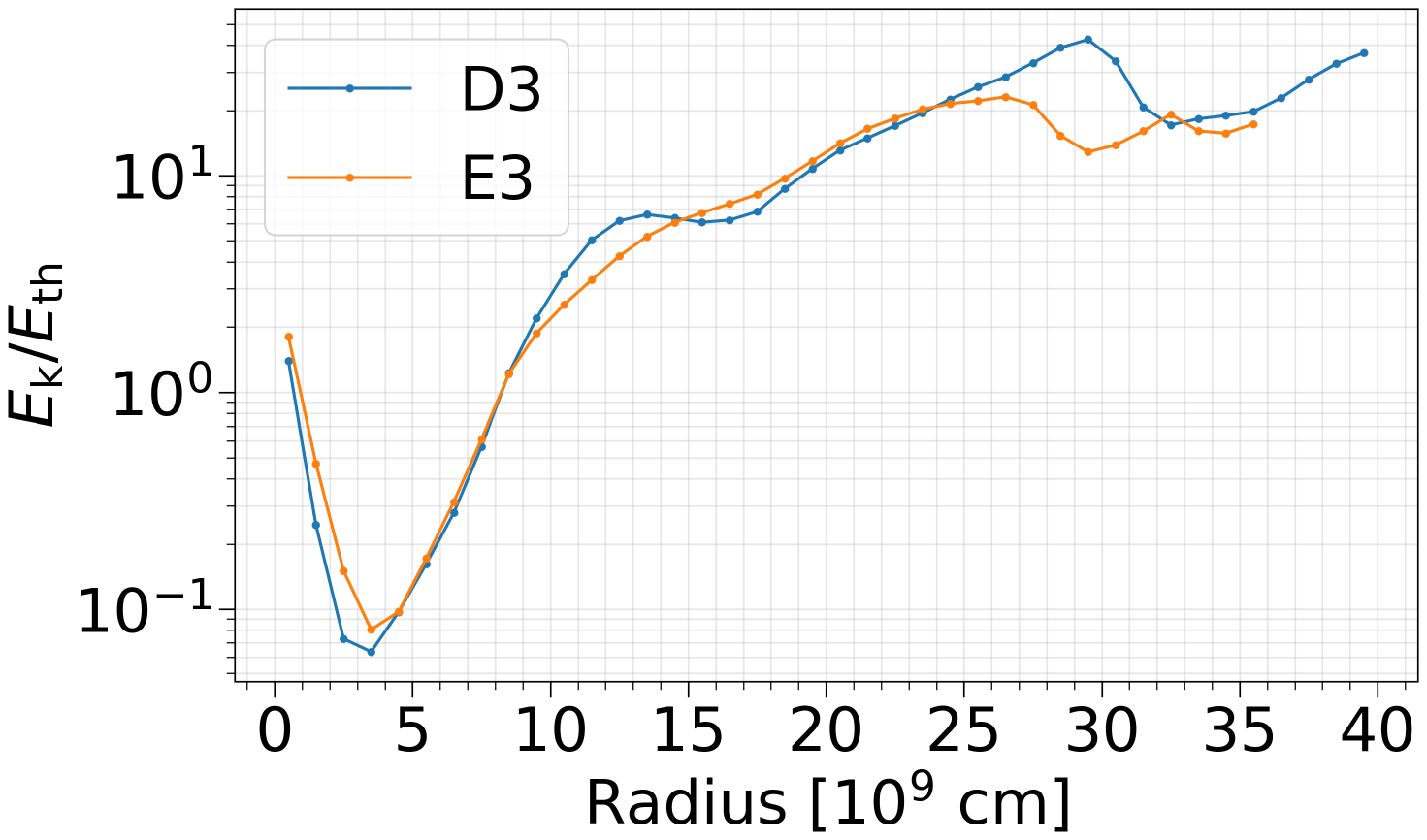} 
\vspace{-0.0cm}
\caption{Radial profile of kinetic-to-thermal energy ratio for simulations D3 (blue) and E3 (orange). Each data point represents a radial interval of $\pm 5\times10^8\cm $ around the plotted radius. The profile extends up to the maximum radius reached by the ejecta at the end of the simulation: $t=11\s$ for simulation D3 and $t=13\s$ for simulation E3.
}
\label{Fig:EnergyRatio}
\end{center}
\end{figure}

In Figure \ref{Fig:EnergyRatio}, we present the ratio of the kinetic to thermal energy, $E_{\rm k}/ E_{\rm th}$,  as a function of radius. The ratio is calculated in spherical shells, each of a width of $10^9 \cm$.   

The ejecta material that is outside the initial radius of the star, $r \gtrsim 8 \times 10^9 \cm$, has larger kinetic energy than thermal energy. This implies that the outer regions, which also expand at much above the escape speed, are already expanding in a homologous outflow. They will maintain their structure until they sweep a large mass of ambient gas. 

The inner ejecta, on the other hand, where $E_{\rm k} \lesssim E_{\rm th}$ did not reach a homologous expansion yet. The study of the late evolution of the inner ejecta is a subject of a future study that will require a much longer simulation time. 

The main conclusion of this section is that although the six jets choke deep inside the core, they manage to imprint a multipolar morphology in the outer ejecta long after shock-breakout.    

\subsection{Filaments and clumps}
\label{subsec:Clumps}

The density maps show not only large lobes and ears, but also much smaller protrusions. Instabilities and vortices form such clumps. The initial perturbations can also result from the asymmetrical pairs of jets. In those cases, the pairs of clumps will form a point-symmetric morphology. We point with pairs of orange arrows at pairs of clumps in the lower right panels of Figure \ref{Fig:densE3}. In many cases, clumps lack point symmetry; these clumps form due to stochastic instabilities.  The evolution of the clumps depends on thermal evolution, namely, heating by shocks and radioactive decay, and radiative cooling. The study of their thermal evolution deserves a separate study. The key result here is that the jets that inflate large bubbles, lobes, and ears can also form pairs of small, dense clumps, and that these clumps need not lie along any jet axis. 

\begin{figure*}
\begin{center}
\vspace{-0.0cm}
\includegraphics[trim=0.1cm 0.0cm 0.0cm 0.0cm ,clip, angle=0, scale=0.55]{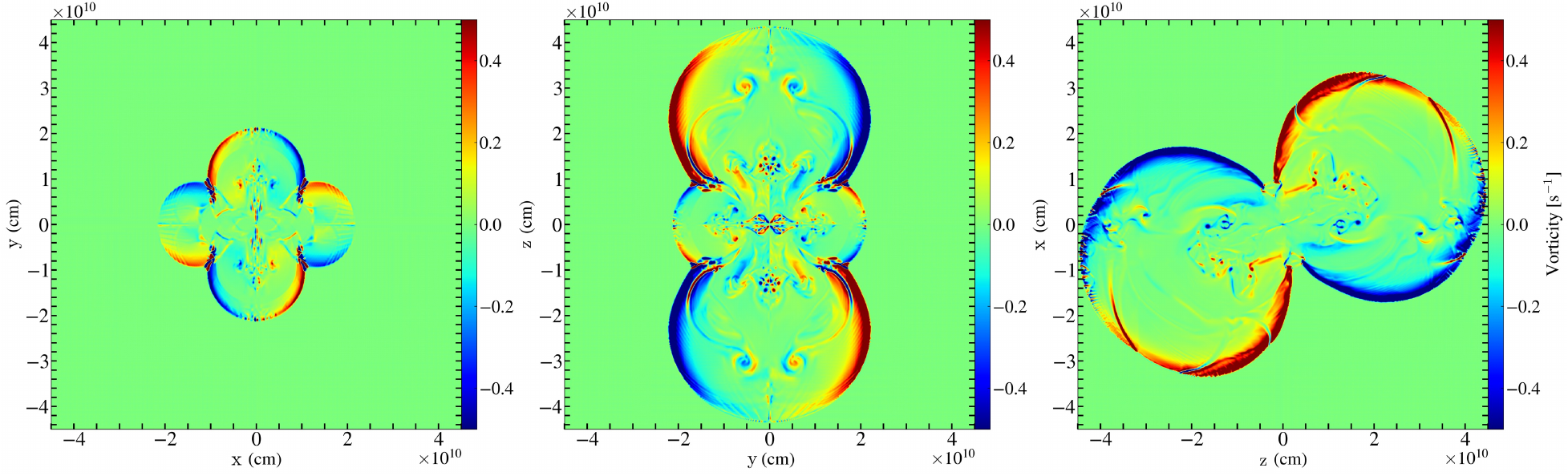}
\vspace{-0.0cm}
\caption{Vorticity maps of simulation E3 at $t=13\s$ in the planes, from left to right, $z=0$, $x=0$, and $y=0$. The maps show the vorticity component perpendicular to the given plane, $(\nabla\times\vec{v})_\perp$. The values are according to the color bar, from $-0.5 \s^{-1}$ (deep blue clockwise) to $0.5 \s^{-1}$ (deep red; counterclockwise). 
}
\label{Fig:VorticityE3}
\end{center}
\end{figure*}

\begin{figure*}
\begin{center}
\vspace{-0.0cm}
\includegraphics[trim=0.1cm 0.0cm 0.0cm 0.0cm ,clip, angle=0, scale=0.55]{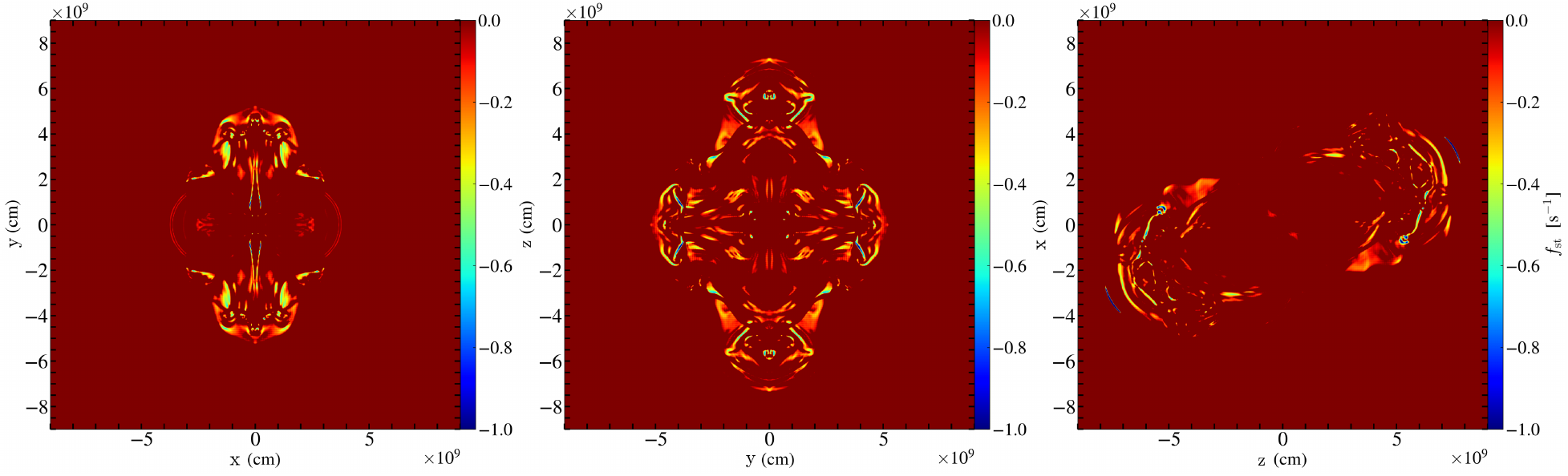} 
\vspace{-0.0cm}
\caption{Rayleigh-Taylor instability (RTI) growth rate maps of simulation E3 at $t=4\s$ in the planes, from left to right, $z=0$, $x=0$, and $y=0$. The panels show the value of $f_{st}$ (equation \ref{eq:rt}) according to the color bars, from $-1 \s^{-1}$ (deep blue) to $\ge 0$ (deep red). Deep-red areas are RTI stable, while others are unstable with a typical growth time of $-1/f_{st}$.
}
\label{Fig:RTE3}
\end{center}
\end{figure*}

We present the vorticity and the growth rate of the Rayleigh-Taylor instability (RTI) for simulation E3 in Figures \ref{Fig:VorticityE3} and \ref{Fig:RTE3}, respectively (the similar figures for simulation D3 are on Zenodo). 
We characterize the vorticity in a plane by the vorticity component perpendicular to that plane, $(\nabla\times\vec{v})_\perp$, and the RTI by its growth rate 
\begin{equation}
f_{st} \equiv \frac{1}{\rho} \sqrt{\lvert \vec{\nabla}P \cdot \vec{\nabla}\rho \rvert} \ \text{sgn}(\vec{\nabla}P\cdot\vec{\nabla}\rho), 
\label{eq:rt}      
\end{equation}
where $\rho$ is the density, $P$ is the pressure, and ${\rm sgn} (\vec{\nabla}P\cdot\vec{\nabla}\rho)$ is the sign of the product of pressure and density gradients. Where $f_{st}$ is positive, the region is stable, while where $f_{st}<0$, the region is unstable with a typical growth rate of $-f_{st}$, and a typical growth time of $-1/f_{st}$. The vorticity maps show regions of strong vorticity on the surfaces of the inflated lobes. The RTI maps show that RTIs are distributed throughout the entire volume; rapidly growing regions (green-blue areas) are found in the outer regions. Instabilities and vortices can form clumps. Due to the stochastic nature of the instabilities, the clumps will not be exactly opposite with respect to the center, but will be closely so because the unstable zones possess point-symmetric structure.   

\begin{figure*}
\begin{center}
\vspace{-0.0cm}
\includegraphics[trim=0.4cm 0.4cm 0.5cm 0.2cm ,clip, angle=0, scale=0.75]{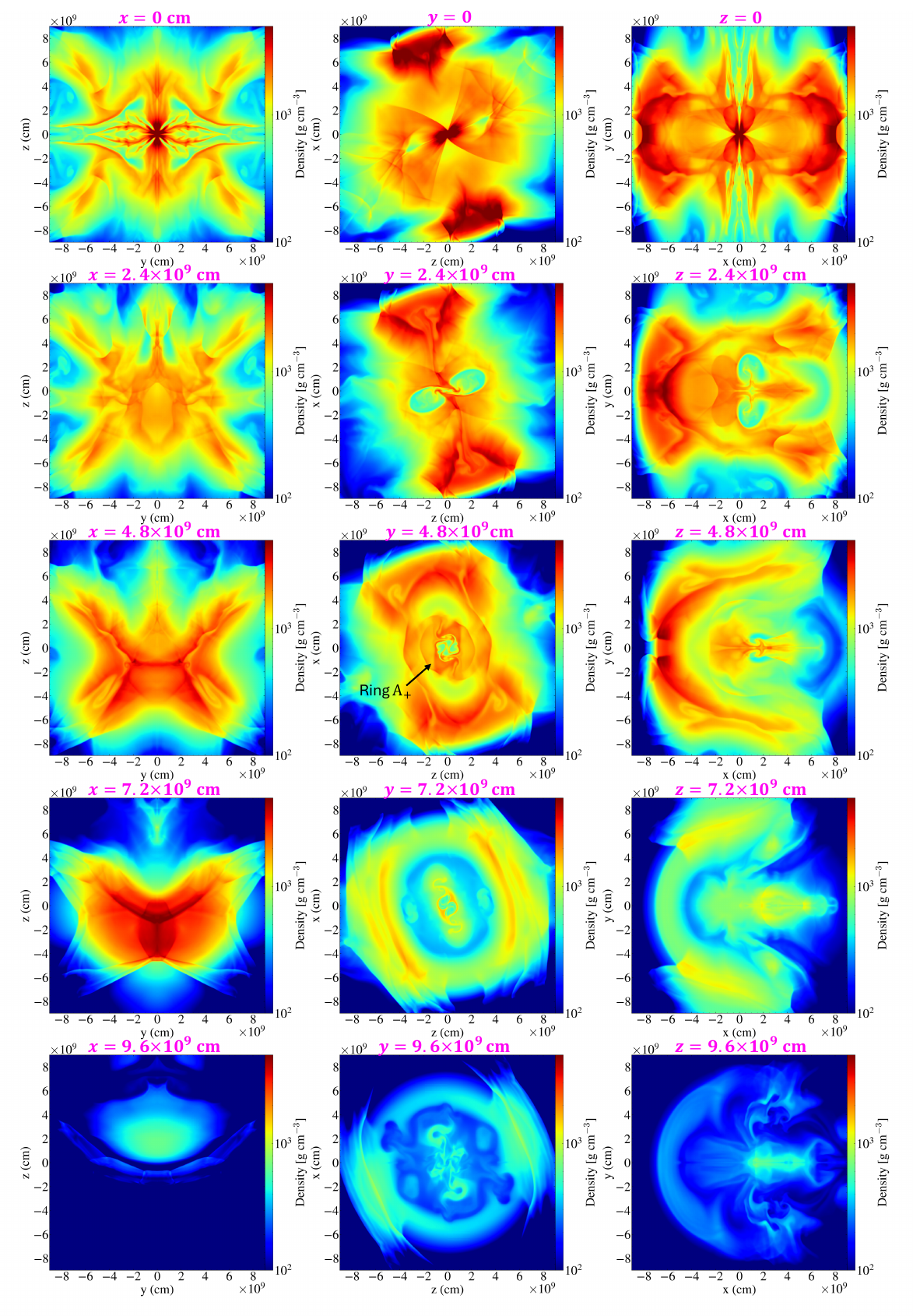} 
\vspace{-0.0cm}
\caption{
Density maps of the inner ejecta for simulation E3 at $t=13 \s$ in several planes parallel to the $x=0$, $y=0$, and $z=0$ planes, as indicated above each panel.   
The density scale is identical across all panels, according to the color bar, from $100 \g \cm^{-1}$ to $5000 \g \cm^{-3}$. Consecutive rows are separated by $2.4 \times 10^{9} \cm$, corresponding to approximately $4\%$ of half the computational domain. The maps reveal a highly asymmetric internal structure, including cavities, arc-like features, and pipe-shaped morphologies whose appearance changes substantially with depth and viewing direction.
}
\label{Fig:DensLayersInnerEqual}
\end{center}
\end{figure*}

To emphasize the clumps and narrow filaments that the jets and instabilities shape in the ejecta, in Figure \ref{Fig:DensLayersInnerEqual} we present the density maps at several planes parallel to the $x=0$, $y=0$, and $z=0$ planes, as indicated above each panel. We focus on the inner ejecta that contains the densest regions. The density maps of the different planets reveal filaments and clumps.  

In \cite{Braudoetal2025}, we present the formation of point-symmetric clumps at early times. Here we follow their evolution for much longer times, long after the shock broke out from the star. The main point of this subsection is that the three pairs of jets can form delicate structures on scales much smaller than the size of the three pairs of bubbles, i.e., narrow filaments and small clumps, that globally might possess point-symmetric structure.
Several CCSNRs possess point-symmetric clump morphologies that studies attributed to the JJEM, including  SNR 0540-69.3 \citep{Soker2022SNR0540}, the Vela SNR (\citealt{SokerShishkin2025Vela}), and Cassiopeia A (\citealt{BearSoker2025}.

\subsection{A blowout}
\label{subsec:Blowout}

\begin{figure}
\begin{center}
\vspace{-0.0cm}
\includegraphics[trim=1.7cm 0.0cm 1.0cm 0.0cm ,clip, angle=0, scale=0.40]{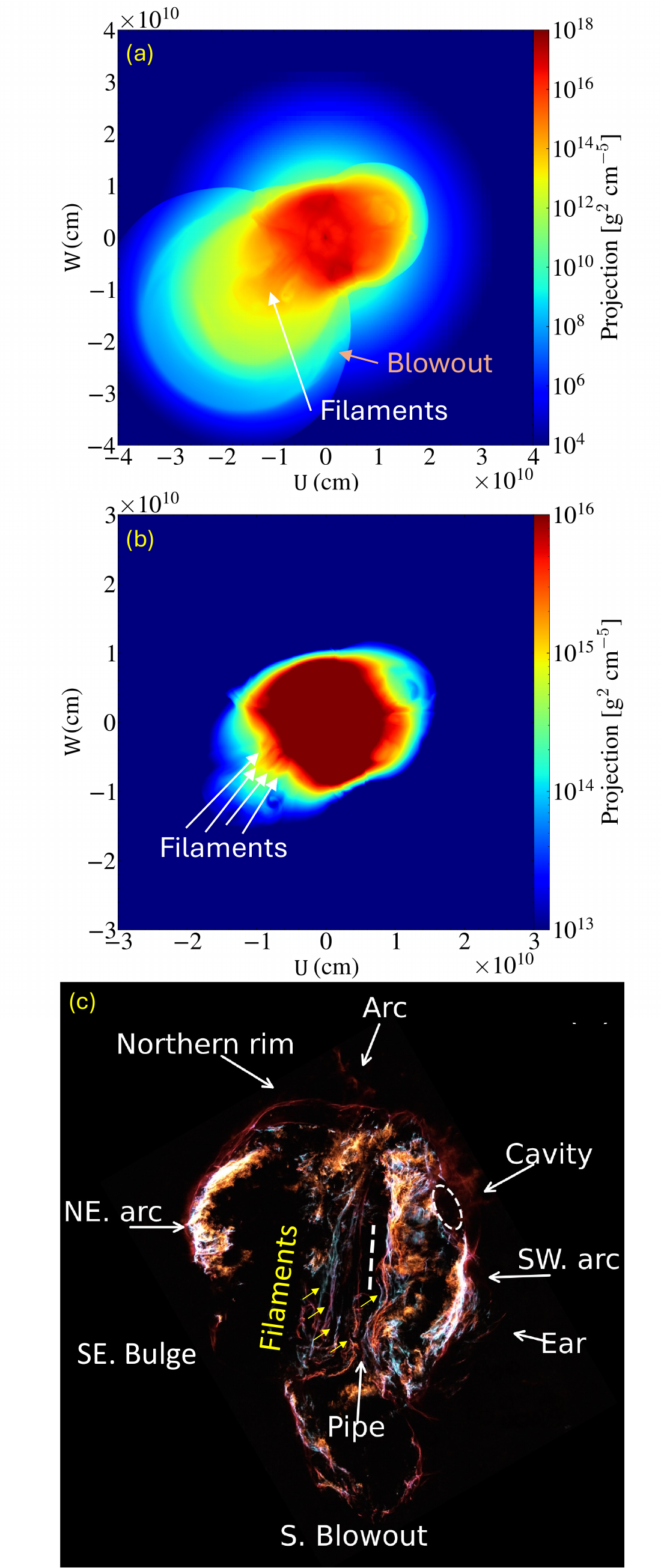} 
\vspace{-0.0cm}
\caption{Blowout and filaments in the simulation D3 and the Cygnus Loop. (a) Numerical emission integral (equation \ref{eq:proj}) of simulation D3, where the integration is along the $y$-axis. The sky coordinates $\rm{(U,W)}$ coincide with the grid coordinates (z,x), respectively. The scale is logarithmic, as indicated by the color bar. (b) Similar to (a), but covering a smaller area and focusing on a narrower emission integral range according to the color bar to better reveal the filaments we point at. (c) A visible image of the Cygnus Loop adapted from \cite{Raymondetal2023}, with white marks from \cite{ShishkinKayeSoker2024}. We mark some filaments with yellow arrows. The simulation qualitatively reproduces the blowout and the filaments leading to it.   
}
\label{Fig:CygnusLoop}
\end{center}
\end{figure}

The most energetic jet in simulation D3, that of energy $E_{\rm k,j} = 3.75 \times 10^{50} \erg$ and an initial direction of $(x,y,z)=(-0.574, 0, -0.819)$, shapes a large protrusion termed a blowout; it is the lower-left part of the ejecta seen in the lower row of Figure \ref{Fig:densD3}.
In Figure \ref{Fig:CygnusLoop} we present the numerical emission integral of simulation D3 integrated along the $y$ axis. 
The numerical emission integral is 
\begin{equation}
    {\rm EI}{\rm(U,W)} = \int \rho^2(x,y,z) \, dL,
    \label{eq:proj}
\end{equation}
where $\rho(x,y,z)$ is the density, ${\rm(U,W)}$ are the coordinates on the plane of the sky perpendicular to the line of sight, and the integration is along the line of sight $L$. The emission integral mimics the observation of optically thin nebulae when the emissivity is proportional to the density squared. In Figure \ref{Fig:CygnusLoop} the integration is along the $y$-axis, so that the coordinates on the plane of the sky ${\rm(U,W)}$ coincide with $(z,x)$. To reveal the full structure, we present the emission integral at two logarithmic scales: from very low to very high intensities, and one that is saturated at low- and high-intensities. 
In the lower panel of Figure \ref{Fig:CygnusLoop}, we present an image of the Cygnus Loop adapted from \cite{Raymondetal2023}, with marks from \cite{ShishkinKayeSoker2024}.  

We find that simulation D3 qualitatively reproduces the south blowout of the Cygnus Loop and the filaments leading to it. The simulation, as is, does not reproduce the bright rim surrounding the blowout in the Cygnus Loop. We attribute the rim to the interaction of the low-density blowout in the Cygnus Loop with an ambient gas, the circumstellar or the interstellar medium. The blowout compresses the ambient gas to form the rim. Our simulation does not extend into the ambient gas (hence, it is not included). 
As we emphasized above, despite the jet in simulation D3 being active only to $t=0.1 \s$ and choking deep in the core, it still managed to shape the blowout and the filaments that extend to it. The lower row of Figure \ref{Fig:densD3} also shows the filaments that lead to the blowout. In the Cygnus Loop, the filaments are much more prominent in the south than in the north. \cite{ShishkinKayeSoker2024} suggests the arc in the very north of the Cygnus Loop might represent a north blowout. They also identified three symmetry axes, which they attributed to three energetic pairs of jets that participated in the explosion of the Cygnus Loop in the framework of the JJEM: One pair of jets with its axis along the pipe where the southern jet in this pair inflated the blowout, a second pair of jets with its axis from the cavity (which is a ring;  \citealt{SokerAkashi2025}) to the SE bulge, and the third with the axis from the SW arc to the NE arc. The point-symmetric morphology of the Cygnus Loop, with several jet-shaped features, strongly supports the JJEM for this CCSNR.     

Overall, our simulation D3 strongly supports the claim by \cite{ShishkinKayeSoker2024} that a jet inflated the blowout in the Cygnus Loop. We further show that the filaments are related to the formation process of the blowout by the jet.

\subsection{The inner ejecta}
\label{subsec:InnerEjecta}

Figure \ref{Fig:EnergyRatio} shows that the thermal energy of the inner ejecta at $r \lesssim 8 \times 10^9 \cm$, about the initial radius of the star, is larger than its kinetic energy. Therefore, this material is not in its homologous expansion, and its exact structure will change until it reaches homologous expansion. The study of its later expansion is a subject of a future study that will demand more computational resources. Here, we highlight some morphological features of this dense material, and in Section \ref{subsec:Doppler} we discuss the Doppler-shift maps. Although the exact structures we discuss below will change, their qualitative structures will remain the same.  

\begin{figure*}
\begin{center}
\vspace{-0.0cm}
\includegraphics[trim=0.5cm 0.4cm 0.0cm 0.0cm ,clip, angle=0, scale=0.95]{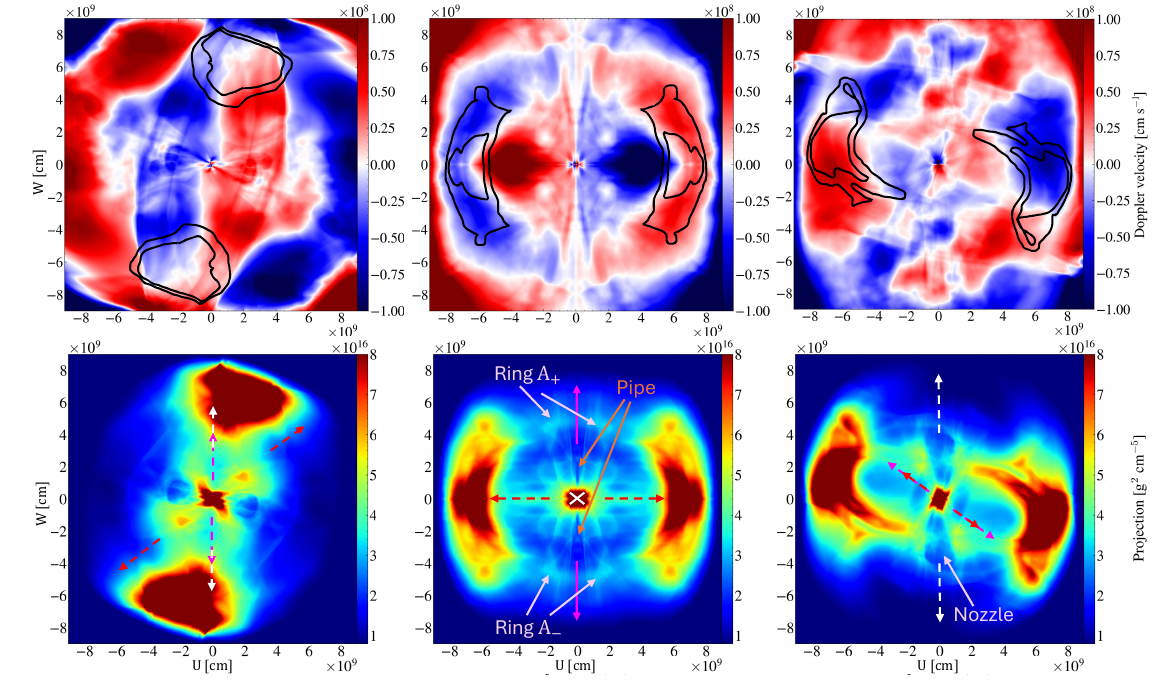} 
\vspace{-0.0cm}
\caption{
Doppler velocity maps and emission integral for simulation E3 at $t=13 \s$, shown for three different viewing directions and for the inner region of the computational domain. Upper panels: Doppler velocity maps (Section \ref{subsec:Doppler}). The color scale spans line-of-sight velocities from $-10^{8} \cm \s^{-1}=-1000 \km \s^{-1}$ to $10^{8} \cm \s^{-1}=1000 \km \s^{-1}$, with blue corresponding to approaching material and red to receding material. Lower panels: Emission integral (equation \ref{eq:proj}) maps. The color scale spans values from $8 \times 10^{15}$ to $8 \times 10^{16} \g^{2} \cm^{-5}$, from deep blue to deep red. Left panels: line-of-sight direction $(\theta,\phi) = (65^{\circ},270^{\circ})$. Central panels: line-of-sight direction $(\theta,\phi) = (0^{\circ},0^{\circ})$, corresponding to a view along the $z$-axis. Right panels: line-of-sight direction $(\theta,\phi) = (45^{\circ},45^{\circ})$. The black contours in the upper panels outline two high-emission integral regions identified in the lower panels. The two contours are for the value of $6 \times 10^{16} \g^{2} \cm^{-5}$ (outer contour) and $8 \times 10^{16}$ (inner contour). 
}
\label{Fig:DopplerInnerE3}
\end{center}
\end{figure*}

The lower panels of Figure \ref{Fig:DopplerInnerE3} show the emission integral at three different angles. 
A jet can leave a faint, narrow, and radially extending, sometimes conical, zone along its propagation direction. The structure of the two opposing marrow-faint regions is called a pipe. We point at a pipe in the middle-lower panel of Figure \ref{Fig:DopplerInnerE3}. The pipe was formed by the last pair of jets in simulation E3, the pair along the $y$-axis. \cite{BraudoSoker2026} compare the pipe morphology in the CCSNRs Cygnus Loop and SNR G292.0+1.8 to simulation E3 and to several jet-shaped planetary nebulae. We will therefore not discuss this structure further. 

In the middle lower panel of Figure \ref{Fig:DopplerInnerE3}, we mark a pair of opposite rings: ${\rm A}_+$ and ${\rm A}_-$. These are thick circum-jet rings whose axis is along the pipe, which is the axis of the jets marked with vertical pink arrows; namely, the plane of each ring is perpendicular to the page. The arrows point at the two places where the rings are bright because the emission integral is integrated through the sides of the ring.  The ring ${\rm A}_+$ is prominent in the plane $y=4.8 \times 10^9$ in Figure \ref{Fig:DensLayersInnerEqual}; we mark this ring there. In the right lower panel of Figure \ref{Fig:DopplerInnerE3}, we mark a pair of opposite 
nozzles; these nozzles have a wider opening than the two sides of the pipe. The axis of the pipe is at directions $(x,y,z)=(0, \pm 1, 0)$ (pink arrows in the middle lower panel), i.,e., along the $y$-axis along the third pair of jets (Table \ref{Tab:Table2}). That of the nozzles is at $(x,y,z)=( 0, 0, \pm 1)$, i.e., the second pair of jets in simulation E3. The axis of the rings and the axis of the nozzles are perpendicular to each other. 

\begin{figure}[]
	\begin{center}
\includegraphics[trim=11.0cm 18.4cm 0.0cm 0.0cm ,clip, scale=0.85]{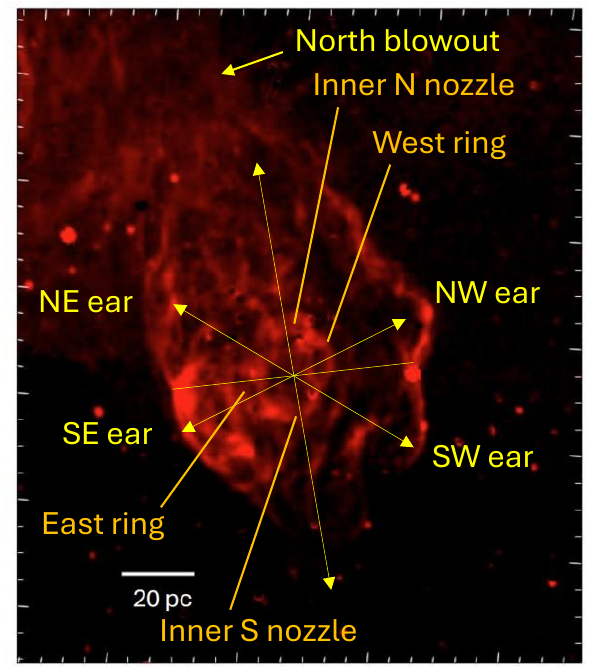} 
\caption{A figure adapted from \cite{Soker2026J0450} who added the marks on an H$\alpha$ emission images of SNR J0450.4-7050 from \cite{Smeatonetal2025SerAJ}. \cite{Soker2026J0450}  identified the structural features and three symmetry axes marked by the double-sided arrows. We argue that the inner ejecta of simulation E3, as we present in the lower panels of Figure \ref{Fig:DopplerInnerE3}, qualitatively reproduce the pair of nozzles and pair of rings in the inner ejecta of SNR J0450.4-7050. We mark the simulated pair of rings in the lower-middle panel of Figure \ref{Fig:DopplerInnerE3}, and the pair of nozzles in the lower-right panel; both pairs are along jet axes.      
}
\label{Fig:SNRJ045047050}
\end{center}
\end{figure}

\begin{figure}
\centering
\includegraphics[trim=0.7cm 1.0cm 1.7cm 0.6cm ,clip, scale=0.45]{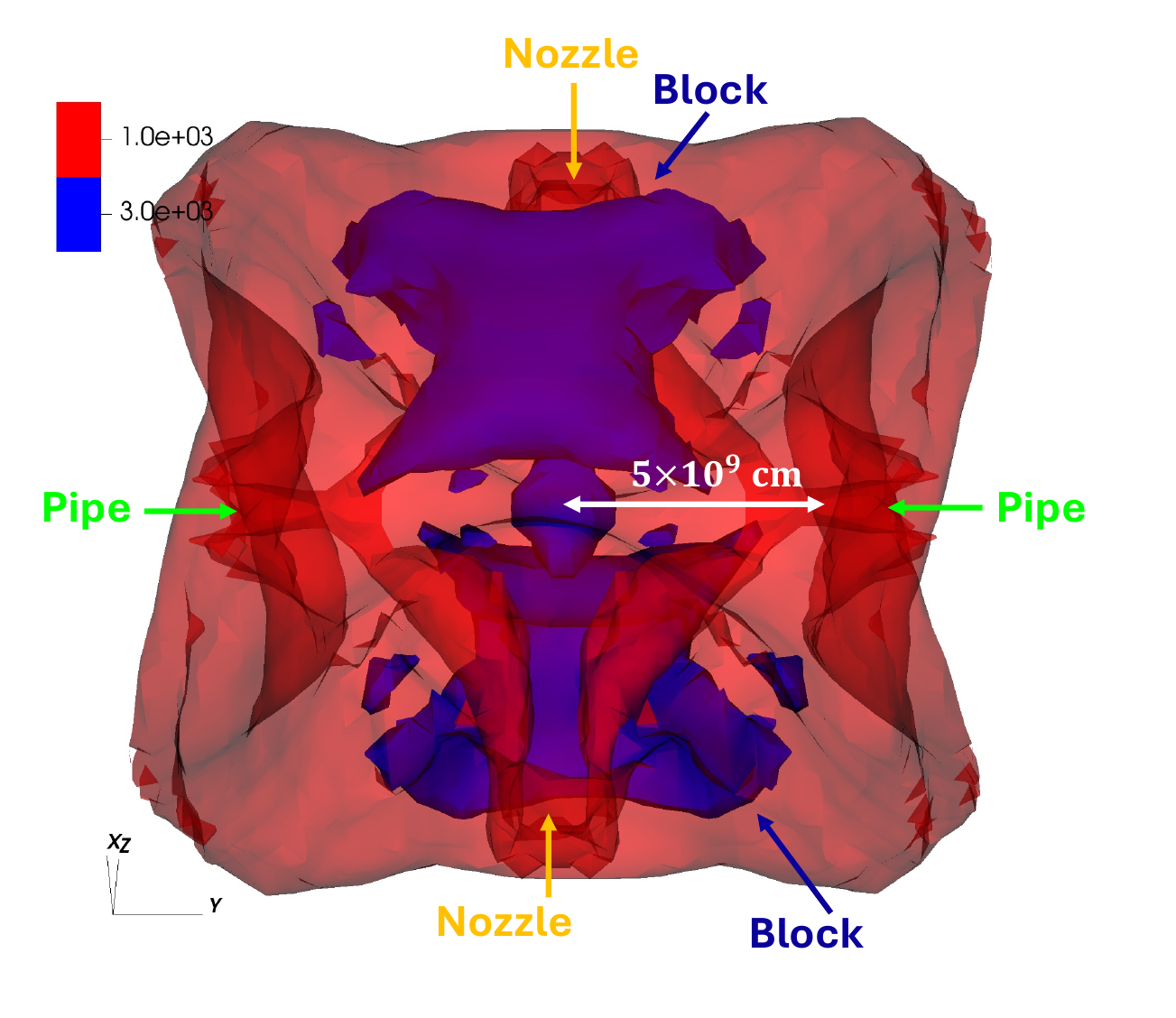} 
\caption{Three-dimensional visualization of the gas density of the inner ejecta in simulation E3. The image is composed of two equidensity semi-transparent surfaces: $1000 \g \cm^{-3}$ (red) and $3000 \g \cm^{-3}$ (blue).  The rings are the dense material surrounding the pipe; they are not narrow but extend for some distance along the pipe. This image emphasizes the two perpendicular axes: the axis of the pipe and its two rings, and the axis of the nozzles (see Figure \ref{Fig:DopplerInnerE3}). This morphology resembles the inner ejecta of SNR J0450.4-7050 in Figure \ref{Fig:SNRJ045047050}.   }
\label{fig:3DVisualization}
\end{figure}

\cite{Soker2026J0450} identified a point-symmetric morphology in the SNR J0450.4-7050, which he attributed to an explosion that includes three pairs of energetic jets in the framework of the JJEM. We present this identification in Figure \ref{Fig:SNRJ045047050}. The inner ejecta of SNR J0450.4-7050 exhibit a pair of rings and a pair of nozzles, which define two of the three symmetry axes. We find that the inner ejecta of simulation E3, which also exhibit a pair of rings and a pair of nozzles along two of the three jets' axes, qualitatively resembles the inner ejecta of SNR J0450.4-7050. 
We cannot present the pair of rings and the pair of nozzles in a single emission-integration image along the $x$-axis because the dense blocks will lie along the line of sight and saturate the image. We present the nozzles, pipe, and blocks in Figure \ref{fig:3DVisualization}, which shows a 3D image composed of two equidensity semi-transparent surfaces. The axis of the two nozzles and the axis of the pipe with its two rings are perpendicular to each other. The two dense blocks appear in this image as well. Again, the structure of the two perpendicular axes resembles that of SNR J0450.4-7050, but not the two dense blocks (blue regions). We expect that launching late jets along the block axis would disperse them, but would require more computing resources (hence, left for a future study).

Overall, we strengthen the claim that three pairs of energetic jets participated in the explosion of SNR J0450.4-7050, in the framework of the JJEM. 

\subsection{Bipolar Doppler maps}
\label{subsec:Doppler}

The upper row of Figure \ref{Fig:DopplerInnerE3} presents Doppler velocity maps. We calculated these maps as follows.
For a given line of sight, specified by angles $(\theta,\phi)$, we calculate the density-weighted projected velocity:
\begin{equation}
v_{\rm D}\rm{(U,W)} = \frac{ \int \rho(x,y,z)\, v_{\rm los}(x,y,z)\, dL}{\int \rho(x,y,z)\,dL},
    \label{eq:dopp}
\end{equation}
where $\rho(x,y,z)$ is the density, $v_{\rm los}(x,y,z)$ is the velocity component along the line of sight, $\rm{(U,W)}$ are the coordinates on the plane of the sky perpendicular to the line of sight, and the integration is along the line of sight $L$. 
Negative values correspond to approaching gas (blueshift) and positive values correspond to receding gas (redshift).

Figure \ref{Fig:DopplerInnerE3} reveals a very interesting property of the inner ejecta, namely, the dense parts, where heavy metals ejected from the core are expected to be concentrated. We refer to the two panels in the middle and the two on the right. The images in the lower panels show a structure around a jet axis: around the third pair of jets along the $y$-axis, depicted by the pink arrows in the lower middle panel, and around the second pair of jets along the $z$-axis, depicted by the dashed white arrows in the lower right panel. If there were only one pair of jets, the morphology would have been axially symmetric. However, the three pairs of jets shape the very dense regions into two opposite blocks (deep red to brown). The Doppler maps in the middle and right upper panels of Figure \ref{Fig:DopplerInnerE3} show that the blocks form a bipolar outflow in the Doppler maps: depending on the viewing angle, one block might be mainly redshifted and one mainly blueshifted.  Namely, the axis of the bipolar structure revealed by the Doppler maps can be highly inclined relative to the axis revealed by the morphology on the plane of the sky. This results from the shaping of the ejecta with three pairs of energetic jets (or more). 

It is worth noting that, while the axis inferred from the morphology in the plane of the sky was formed by a pair of jets, the axis connecting the two very dense Doppler-shifted blocks is not. The jets form the two opposite blocks in regions between the jet axes: the densest parts in the two very dense blocks are not along any jet axis, but rather between the axes we draw on the lower row panels of Figure \ref{Fig:DopplerInnerE3}. This is similar to the formation of a dense bar from one side to the ejecta to the other between jets' axes that \cite{AkashiSoker2026B} obtained with two pairs of energetic jets.

With this structure of a morphological axis on the plane of the sky that is highly inclined to a Doppler bipolar outflow, we examine the Doppler maps and images of two SNRs and deduce that two, and more likely, three or more pairs of jets shaped their structures. This strongly supports the claim that the JJEM explains the explosion of their progenitors. 

\begin{figure}[]
	\begin{center}
\includegraphics[trim=0.0cm 9.4cm 0.0cm 0.0cm ,clip, scale=0.4]{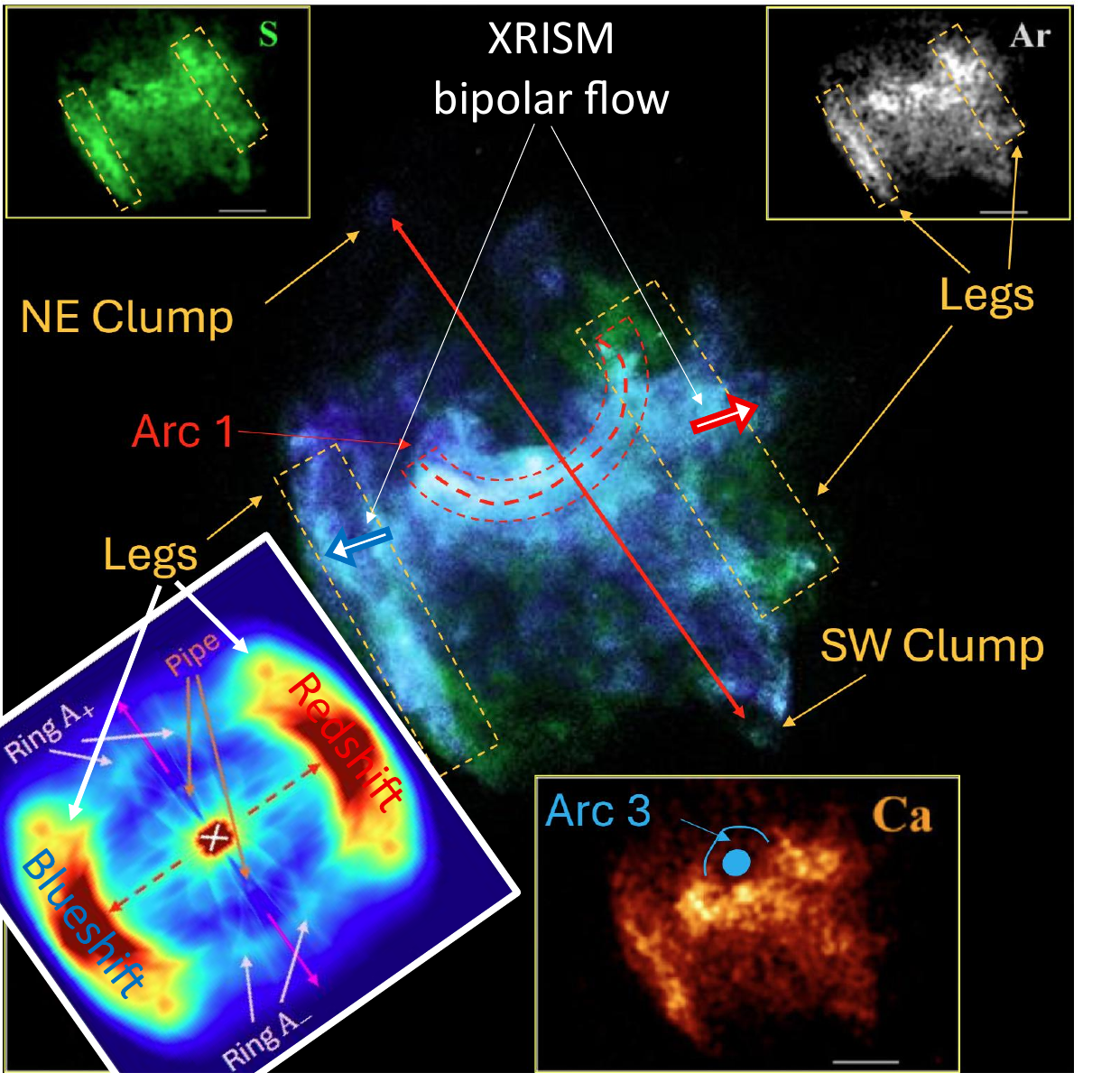} 
\caption{A figure adapted from \cite{SokerShishkin2025W49} who made the marks of morphological features on an X-ray image from Chandra ({\url{https://chandra.harvard.edu/photo/2013/w49b/}}; credit NASA/CXC/MIT/\citealt{Lopezetal2013a}). The three inset images are from \citet{Lopezetal2013a} with the scale bar marking $1^\prime$. The double-sided red arrow is the main jet axis they identified. We added the marks in white, the marks of the redshifted and blueshifted direction based on \cite{XRISMW49B2025}, and the lower-left inset, which is the rotated middle-lower panel from Figure \ref{Fig:DopplerInnerE3}. Our simulation reproduces the high inclination of the main jet axis and the Doppler bipolar outflow.  
}
\label{Fig:W49B}
\end{center}
\end{figure}

In Figure \ref{Fig:W49B} we present X-ray images of SNR W49B, adapted from \cite{SokerShishkin2025W49b}, who marked the morphological features, including the main jet axis, by the double-sided red arrow. \cite{BearSoker2017} and \cite{SokerShishkin2025W49b} defined this main jet axis based on the morphology of SNR W49B.  \cite{XRISMW49B2025} analyzed the velocity structure of W49B as they inferred from XRISM observations, and identified a bipolar outflow with Doppler-shifted velocities from $\simeq - 300 \km \s^{-1}$ to $\simeq +300 \km \s^{-1}$. We mark the projection of the Doppler-shifted outflow on the plane of the sky with the two white-filled arrows. The Doppler-shifted bipolar axis is the same as some earlier claims (e.g., \citealt{Keohaneetal2007, Lopezetal2013a}). 
We here resolve this dispute between the claim for almost perpendicular symmetry axes. Both claims are correct. Simply, the explosion by three or more pairs of energetic jets can form both a morphological symmetry axis and a highly inclined Doppler-shifted bipolar outflow of denser material. 
We added the lower-left inset in Figure \ref{Fig:W49B} and rotated it to emphasize the qualitative similarity of our results to the observations of SNR W49B. 

We here resolve the dispute over the bipolar axis of SNR W49B by showing that an explosion by three pairs of energetic jets can form a symmetry axis along one or more jet axes, together with a bipolar outflow of the densest gas that reveals a distinct bipolar axis in Doppler maps. The densest gas blocks are not along a jet axis, but rather were compressed between the jet axes.  The key is that three or more pairs of jets exploded the progenitor of SNR W49B, as predicted by the JJEM. 

\begin{figure}[]
	\begin{center}
\includegraphics[trim=0.0cm 2.4cm 0.0cm 0.0cm ,clip, scale=0.4]{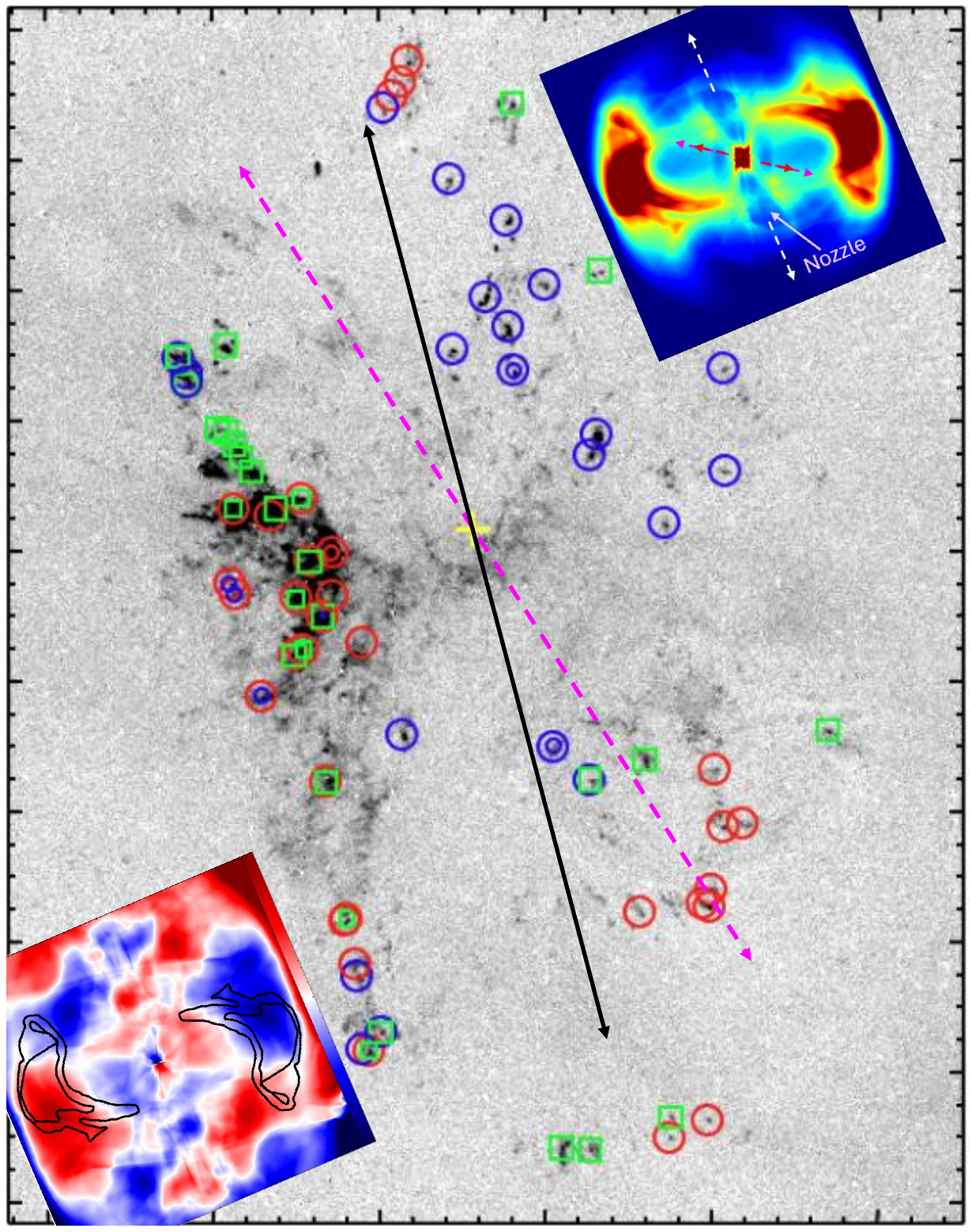} 
\caption{An image of SNR G292.0+1.8 adapted from \cite{Plunkettetal2026} of a continuum-subtracted [O \textsc{iii}] emission. Colored circles indicate Doppler-shifted velocities:  
red: $v_{\rm rad} > 330 \km \s^{-1}$; blue: $v_{\rm rad} < - 330 \km \s^{-1}$; green: $-330 \km \s^{-1} < v_{\rm rad} < 330 \km \s^{-1}$. 
The yellow cross is the expansion center from \cite{Winkleretal2009}. The black solid double-sided arrow is the symmetry axis that \cite{Bearetal2017} defined by the two opposite ears it connects, and the pink dashed double-sided arrow is a possible symmetry axis that they defined by two bright H$\alpha$ arcs in an image from \cite{Ghavamianetal2005}.
The insets are the two right panels of Figure \ref{Fig:DopplerInnerE3}, demonstrating that we can reproduce the highly inclined symmetry axis (dashed white line in the upper inset) to the Doppler bipolar outflow (lower inset). 
}
\label{Fig:G292}
\end{center}
\end{figure}

Another SNR with a Doppler bipolar outflow inclined to morphological axes is SNR G292.0+1.8. 
In Figure \ref{Fig:G292} we present an [O \textsc{iii}] emission image adapted from \cite{Plunkettetal2026}, who also mark knots with redshifted (red circles) and blueshifted (blue circles) velocities. With two double-sided arrows, we added two symmetry axes from \cite{Bearetal2017}: The black solid double-sided arrow connects two ears that they identified in X-ray images (e.g., by \citealt{Ghavamianetal2012}). The pink dashed double-sided arrow is a possible symmetry axis that \cite{Bearetal2017} defined by the two bright H$\alpha$ arcs in an image from \cite{Ghavamianetal2005}. These two morphological axes define a general morphology axis along a direction between them, which is highly inclined to the general Doppler bipolar outflow.  
We added two insets to Figure \ref{Fig:G292}: the lower-left inset shows a Doppler map of simulation E3, and the upper-right inset shows an emission integral map, as shown in the two right panels of \ref{Fig:DopplerInnerE3}.  

The [O \textsc{iii}] emission image of SNR G292.0+1.8 in Figure \ref{Fig:G292} exhibits a general `H-shaped' structure with a complex Doppler shift structure. The west (right) leg of the H-shaped structure is mostly blueshifted, with redshifted regions at its ends. The east leg of the H-shaped structure is mostly redshifted, with scattered blueshifted knots. The Doppler map does not yield a well-defined bipolar axis, but has a general one from southeast to northwest. The Doppler bipolar outflow is inclined to the axes defined by morphology, and which \cite{Bearetal2017} attributed to a jet axis. We can make a parallel between our simulation E3 and the structure of SNR G292.0+1.8. The pair of jets that shaped the pair of nozzles, i.e., the white-dashed arrows in the upper inset, suggests that, as \cite{Bearetal2017} argued, a pair of jets shaped the ears, and the general faint zone along the two double-sided arrows in Figure \ref{Fig:G292}. The densest parts, deep red and brown in the upper inset, are parallel to the legs of the H-shaped structure of SNR G292.0+1.8. The lower inset shows that the densest parts, marked with black contours, have a complex Doppler-shift structure that is also qualitatively parallel to the observed one, namely, a mixed blue-shifted and redshifted region on each side of the morphological symmetry axis. The right side of the simulation shows a general blueshifted outflow, with redshifted regions at its end, qualitatively similar to the Doppler-shift structure observed in the west leg of SNR G292.0+1.8. 

Overall, our simulation explains the morphology and the Doppler-shift structure of SNR G292.0+1.8. The key is that three or more pairs of jets exploded the progenitor of SNR G292.0+1.8, as predicted by the JJEM. 

\section{Summary}
\label{sec:Summary}

We presented the results of two 3D hydrodynamic simulations of the JJEM, in which we launched three pairs of inclined opposite jets into the core of an enveloped-stripped stellar model. We launched the jets within about a second and followed the ejecta for more than 10 seconds, long after the jets ceased and the shock breakout. In simulation D3, the jets differ from each other in several properties (Table \ref{Tab:Table1}), while in simulation E3, the jets are equal in their properties, besides their directions (Table \ref{Tab:Table2}). 

Our simulations reproduced some morphological features of observed CCSNRs that a single pair of jets or instabilities alone cannot reproduce. We emphasize that we were not aiming at any particular CCSNR in setting the numerical calculations, but rather chose the initial jet properties for general demonstration of some JJEM properties. Therefore, we do not reproduce all properties of the four CCSNRs we analyzed, but we do reproduce some prominent morphological features in each of them. These similarities support the JJEM.   

Our main findings are as follows. 
\begin{enumerate}
    \item Despite the fact that the jets are choked deep inside the star and cease to be active long before shock-breakout, the jets nonetheless manage to form a pronounced multipolar (point-symmetric) morphology, in the outer ejecta, e.g., Figures \ref{Fig:densD3} and \ref{Fig:densE3}, and in the inner ejecta \ref{Fig:DopplerInnerE3}. The kinetic energy of the outer ejecta is much larger than its thermal energy (Figure \ref{Fig:EnergyRatio}), and it is already in an almost homologous expansion; it will maintain structure (Section \ref{subsec:Multipolar}).    
    \item  The interaction of the jets with the stellar material excites vortices (Figure \ref{Fig:VorticityE3}) and is unstable with a short RTI growth time (Figure \ref{Fig:RTE3}). The instabilities and vortices form clumps and filaments; we point out some pairs in Figure \ref{Fig:densE3} (Figure \ref{Fig:DensLayersInnerEqual} shows more filaments in the density maps across different planes). Although the exact clump formation is stochastic due to the nature of instabilities, the unstable zones form a point-symmetric morphology due to the jet-star interaction. Therefore, clumps and filaments can form point-symmetric morphologies as observed in some CCSNRs, e.g.,  SNR 0540-69.3, Vela, and Cassiopeia A (Section \ref{subsec:Clumps}).   
    \item The most energetic jet of simulation D3 breaks out and forms a large low-density blowout, as the lower row of Figure \ref{Fig:densD3} presents. In Figure \ref{Fig:CygnusLoop} (Section \ref{subsec:Blowout}), we present an emission integral from simulation D3 that mimics an observation, and compare it with an image of the Cygnus Loop CCSNR. We find similarities in the blowout and the filaments that lead to it. The Cygnus Loop blowout has a bright boundary. We attribute it to the interaction between the blowout and the ambient gas, i.e., the circumstellar material the progenitor lost prior to the explosion; the blowout compresses the ambient gas to a thin shell. The filaments we find are not as thin and long as the observed ones. It is likely that a higher numerical resolution than we have and the inclusion of radiative cooling will result in longer, narrower filaments. 
    \item The inner ejecta in our simulations has not yet undergone a homologous expansion by the end of the simulation (its thermal energy is still larger than its kinetic energy; Figure \ref{Fig:EnergyRatio}). Its structure will change, but we expect its qualitative structure to persist. We present the inner ejecta in Figure \ref{Fig:DopplerInnerE3}. It exhibits a pair of rings and a pair of nozzles along the axes of two perpendicular pairs of jets. We argued (Section \ref{subsec:InnerEjecta}) that this resembles the inner ejecta of SNR J0450.4-7050 that we present in Figure \ref{Fig:SNRJ045047050}.  
    \item The upper row of Figure \ref{Fig:DopplerInnerE3} presents the Doppler shift maps weighted by the density (equation \ref{eq:dopp}). Comparing these with the emission integral maps in the lower row of Figure \ref{Fig:DopplerInnerE3}, we found that, in simulation E3, the two axes the emission integral shows, that of the rings and nozzles, are at a high angle to the bipolar expansion that the Doppler maps show. With this, we solve the puzzle of two CCSNRs whose bipolar Doppler-shifted outflows are at a large angle to the morphological axis. We present images of these two CCSNRs, W49B (Figure \ref{Fig:W49B}) and SNR G292.0+1.8 (Figure \ref{Fig:G292}), and compare them to the numerical results in the insets of these figures. The key is that three or more energetic pairs of jets exploded these CCSNRs. Our arguments for three or more pairs of jets in W49B support the claim that it is a CCSNR. The high iron abundance in W49B can result from vigorous nuclear burning (thermonuclear outburst) of a rare mixed helium-oxygen layer, as \cite{Klimovetal2026} suggested for CCSNRs W49B and 3C 397. We differ from them in claiming three pairs of jets rather than just one.      
\end{enumerate}

To summarize, by showing that the jets can shape multipolar morphologies even if they are choked deep inside the star, and by reproducing structures as observed in CCSNRs, particularly the four CCSNRs we analyzed here, our study adds significant support to the claim that jittering jets (the JJEM) explode most, and maybe all, CCSNRs. 
 
\section*{Acknowledgements}


  \bibliography{bib.bib}

\end{document}